\documentclass[twocolumn,showpacs,amsmath,amssymb,prl]{revtex4}

\usepackage[colorlinks=true,linkcolor=red, linktocpage=true]{hyperref}

\usepackage{graphicx,amsmath,amssymb,bm,multirow}
\usepackage{amsthm}
\usepackage{amsfonts}
\usepackage{dcolumn,CJK}
\usepackage{bm}
\usepackage{mathrsfs}
\usepackage[usenames,dvipsnames]{color}

\begin{document}

\title{Beyond relativistic mean-field approach for nuclear octupole excitations}
 \author{J. M. Yao}
 \affiliation{School of Physical Science and Technology, Southwest University, Chongqing 400715, China}
\affiliation{Department of Physics, Tohoku University, Sendai 980-8578, Japan}

\author{E. F. Zhou}
\affiliation{School of Physical Science and Technology, Southwest University, Chongqing 400715, China}
\author{Z. P. Li}
\affiliation{School of Physical Science and Technology, Southwest University, Chongqing 400715, China}

\date{\today}

\begin{abstract}
We report the first beyond-mean-field study of low-lying parity-doublet states in $^{224}$Ra by extending the multireference relativistic energy density functional method to include dynamical correlations related to symmetry restoration and quadrupole-octupole shape fluctuation with a generator coordinate method combined with parity, particle-number, and angular-momentum projections. We clarify full microscopically that the origin of spin-dependent parity splitting in low-spin states is related to the octupole shape stabilization of positive-parity states, the dominated shapes of which drift gradually to that of negative-parity ones.

\end{abstract}

\maketitle


The existence of octupole shaped nuclei in intrinsic frame has been suggested very early from the observed low-lying parity doublets connected with strong electric odd-multipole ($E\lambda$) transition strengths~\cite{Ahmad93,Butler96}. Among them, $^{224}$Ra was recently suggested to be a stable pear shaped nucleus based on the measured $E\lambda$ transitions~\cite{Gaffney13}. Nevertheless, as pointed out early~\cite{Eisenberg87} and shown in this paper that the effects stemming from static octupole shapes on $E\lambda$ transitions can be almost equally well described by including dynamical octupole deformations around the equilibrium shape. Moreover, the spectroscopy of reflection-asymmetric diatomic molecules suggests the existence of a rotational band with alternating parity, which has indeed been found in some candidates of octupole shaped nuclei at high-spin states. However, an evident energy splitting is observed in the odd and even parity states of these nuclei at low-spin region. In particular, this parity splitting is gradually increased with the decrease of spin. This peculiar feature indicates the possible existence of large quantum shape fluctuation in octupole shapes at low-spin states.   Therefore, whether atomic nuclei possess dynamical or stable octupole shapes in low-spin states remains an open question, which is relevant not only for deepening our understanding on nuclear structure at low energy, but also in searching for CP-violating Schiff moment which is expected to be amplified in octupole deformed nuclei~\cite{Engel13}.

Numerous studies have been performed to search for octupole deformed nuclei with selfconsistent mean-field approaches~\cite{Marcos83,Bonche86,Robledo87,Egido91,Geng07,Zhang10,Guo10}, and to understand nuclear collective octupole excitations with energy density functional (EDF) based collective Hamiltonian~\cite{Egido89,Robledo10,Robledo11,Li13}, EDF mapped interacting boson model~\cite{Nomura14}, and generator coordinate method (GCM)~\cite{Heenen01,Guzman12,Robledo13}. These studies have greatly deepened our understanding on  the structure of octupole candidate nuclei. However, the onset of unusual spin-dependent parity splitting has not been clarified. The combination of energy density functional approach with GCM provides a powerful tool to study nuclear shape fluctuations. Unfortunately, angular-momentum projection has not yet been implemented in this framework for octupole nuclei and the stabilization of nuclear octupole shape against rotation remains to be examined.

In the meantime, cranked mean-field approaches have also been adopted to study nuclear octupole deformation in different spin states~\cite{Nazarewicz84,Garrote97}. A spin-dependent staggering in deformations and the energy splitting of odd and even parity states were observed in the $N=88$ isotones~\cite{Garrote97}, where a phase transition picture from octupole vibration to octupole deformation was suggested. However, the effects of shape fluctuation and angular momentum conservation remain to be taken into account. On the other hand, to explain the spin-dependent parity splitting, Jolos et al. proposed a picture of rotation-induced second-order phase transition from dynamical octupole to static octupole shapes based on a one-dimensional collective model with phenomenological spin-dependent potentials~\cite{Jolos94}, which seems to be consistent with semiclassical analysis of available data~\cite{Ahmad93,Butler96,Cocks97,Wiedenhover99}. In contrast, Shneidman et al. proposed a dynamical cluster picture with oscillations in mass asymmetry coordinate based on a phenomenological dinuclear model~\cite{Shneidman02}. In these phenomenological studies, however, the E$\lambda$ transition properties were seldom examined and the microscopic foundation of their model assumptions was not clear. In view of all these facts, it is important and timely to establish a full microscopic approach to study possible octupole shaped nuclei with alternating parity states and to unveil the underlying mechanism responsible for the spin-dependent parity splitting at low-spin region.

%


To this end, we extend the selfconsistent relativistic energy density functional method~\cite{Ring96,Vretenar05,Meng06} to include dynamical correlations related to symmetry restoration and quadrupole-octupole shape fluctuation with the GCM combined with  parity, particle-number, and angular-momentum  projections. The symmetry conserved wave function for the low-lying collective states is constructed by superposing a set of quantum-number projected nonorthogonal mean-field reference states $\vert q\rangle$ around the equilibrium shape
\begin{equation}\label{gcmwf}
\vert J^\pi NZ; \alpha\rangle
=\sum_{\kappa\in\{q, K\}} f^{J\pi\alpha}_\kappa  \hat P^J_{MK} \hat P^N\hat P^Z  \hat P^\pi\vert q\rangle,
\end{equation}
where the generator coordinate $q$ stands for the discritized deformation parameters $\{\beta_2,  \beta_3\}$ of the reference states from deformation constrained selfconsistent mean-field calculation based on a universal relativistic energy functional PC-PK1~\cite{Zhao10}.  The $\hat P^G$s ($G\equiv J, \pi, N, Z )$ are projection operators~\cite{Ring80}. The weight function $f^{J\pi\alpha}_\kappa$ is determined by the Hill-Wheeler-Griffin equation,
\begin{equation}\label{HWG}
\sum_{\kappa_b} \left[  \mathscr{H}^{J\pi}_{\kappa_a, \kappa_b}-E^{J\pi}_{\alpha} \mathscr{N}^{J\pi}_{\kappa_a, \kappa_b}\right]f^{J\pi\alpha}_{\kappa_b}=0
\end{equation}
where the hamiltonian kernel $ \mathscr{H}^{J\pi}_{\kappa_a, \kappa_b}$ and norm kernel
$\mathscr{N}^{J\pi}_{\kappa_a, \kappa_b}$ are given by,
\begin{equation}
\label{kernel}
 \mathscr{O}^{J\pi}_{\kappa_a, \kappa_b}
 =\langle q_a \vert  \hat O \hat P^J_{K_aK_b} \hat P^N\hat P^Z  \hat P^\pi\vert q_b\rangle
\end{equation}
with the operator $\hat O$ representing $\hat H$ and $1$, respectively.

In the present work, the mean-field wave functions $\vert q\rangle$ are obtained by selfconsistent deformation constrained calculation, where the Dirac equation of single-particle wave function is solved in a three dimensional isotropic harmonic-oscillator basis with fourteen major shell. Although this basis space is not sufficient  for the absolute energies of states, it provides reasonable convergent solutions to nuclear spectroscopic properties, including excitation energies and transition strengths. Axial and time-reversal symmetries are imposed to reduce the computational burden. Moreover, since the BCS method and Bogoliubov transformation give similar good description for the low-lying states of nuclei not far from stability line if the pairing strengths are chosen properly~\cite{Xiang13}, we adopt the computation less demanded BCS method to take into account the pairing correlations between nucleons with the strength parameters of zero-range pairing forces chosen according to the PC-PK1 parametrization~\cite{Zhao10}. The Pfaffian techniques~\cite{Robledo09,Bertsch12} are implemented to calculate norm overlaps in the norm kernel. The hamiltonian kernel is calculated with mixed-density prescription~\cite{Lacroix09}. The numbers of mesh points in rotation Euler angle and gauge angle are chosen as 16 and 9, respectively, in the angular momentum and particle number projections. After convergence check, we finally choose 39 reference states in the $(\beta_2, \beta_3\ge0)$ deformation plane  in the GCM calculation. The configurations with $\beta_3<0$ are included automatically by using the parity operator.

\begin{figure}[tb]
\centering
\includegraphics[width=8.5cm]{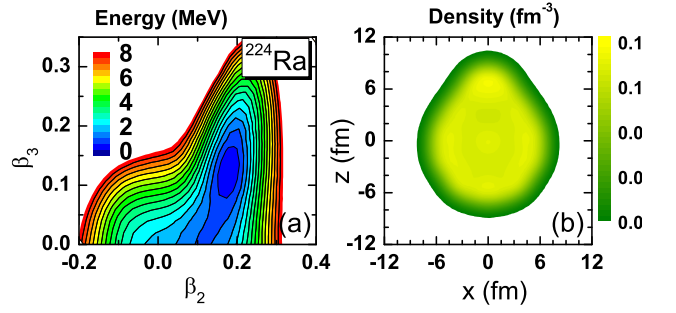}
\caption{\label{fig2} (Color online) (a) Mean-field energy surface of $^{224}$Ra
in $\beta_2$-$\beta_3$ deformation plane (normalized to energy minimum),
where two neighboring lines are separated by 0.5 MeV. (b) Total nucleon density of the energy-minimum state
with $\beta_2=0.179, \beta_3=0.125, \beta_4=0.146$ in $y=0$ plane.}
\end{figure}

It is worth mentioning that the GCM based on nonrelativistic Skyrme~\cite{Bender08} or Gogny~\cite{Rodriguez10} forces has recently been extended for odd-particle states~\cite{Bally14} and cranked states~\cite{Borrajo15}, respectively with great efforts. However, these approaches are presently limited to quadrupole shapes.  The present method is a further extension of our previous beyond mean-field method for triaxial nuclei~\cite{Yao10} by allowing for octupole shape degree of freedom and provides the most advanced beyond mean-field calculation for nuclear octupole excitations.

Figure~\ref{fig2}(a) displays mean-field energy surface of $^{224}$Ra in $\beta_2$-$\beta_3$ deformation plane, where the deformation parameters $\beta_\lambda$ are defined as
\begin{equation}\label{deformation}
  \beta_\lambda \equiv \dfrac{4\pi}{3A R^\lambda}\langle q \vert r^\lambda Y_{\lambda 0}\vert q\rangle, \quad R=1.2A^{1/3},
\end{equation}
with $A$ being mass number of the nucleus. The topography of the mean-field energy surface in Fig.~\ref{fig2} is essentially the same as that obtained in Ref.~\cite{Nomura14} with a relativistic Hartree-Bogoliubov (RHB) model. The global energy minimum is located at the configuration with deformations $\beta_2=0.179, \beta_3=0.125, \beta_4=0.146$, which are slightly larger than the values $\beta_2=0.154, \beta_3=0.097, \beta_4=0.15$ in Ref.~\cite{Gaffney13}. Total nucleon distribution of this energy-minimum configuration is depicted in Fig.~\ref{fig2}(b), exhibiting a well-developed pear shape. The energy difference between the energy minimum and the lowest reflection-symmetric configuration is, however, not large enough to prevent shape fluctuation. The present GCM calculation shows a broad distribution of the collective wave function in $\beta_2$-$\beta_3$ plane for ground state, consistent with the finding in Ref.~\cite{Robledo13}. With the increase of spin, the spread collective wave function of positive-parity states becomes gradually concentrated around the energy minimum and close to that of negative-parity states, which is more stable against rotation.

\begin{figure}[tb]
\centering
\includegraphics[width=9.5cm]{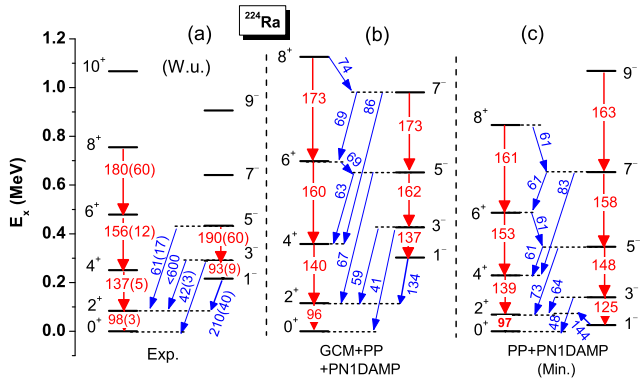}\vspace{-0.5cm}
\caption{\label{fig3} (Color online) Low-lying energy spectra for $^{224}$Ra. The available data from Ref.~\cite{Gaffney13} and the results from full-configuration and the single energy-minimum configuration calculations are shown in (a), (b) and (c), respectively. The numbers on arrows are E2 (red color) and E3 (blue color)
transition strengths (Weisskopf units). }
\end{figure}

\begin{figure}[tb]
\centering
\includegraphics[width=5.5cm]{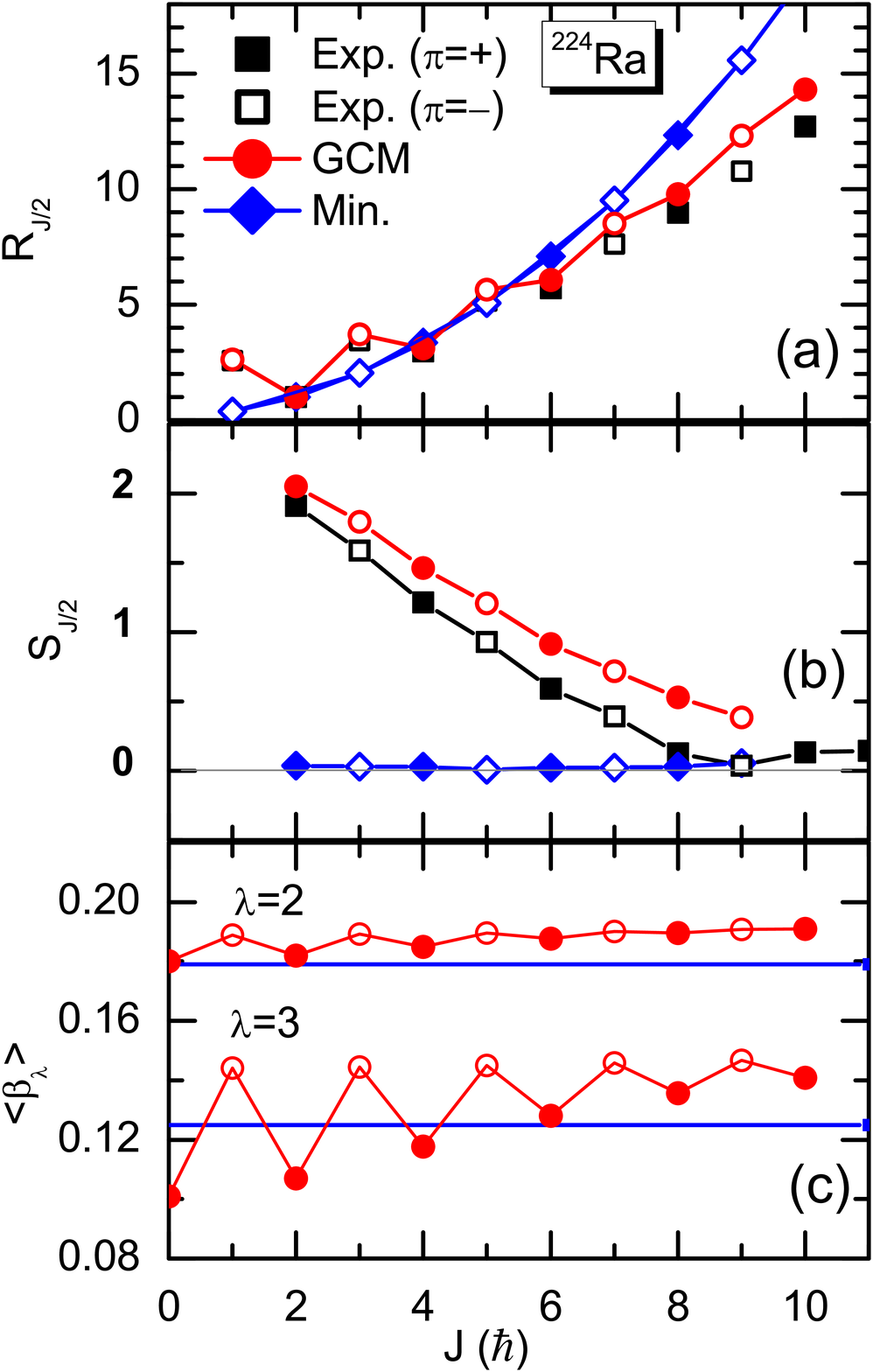}\vspace{-0.5cm}
\caption{\label{fig4} (Color online) (a) Excitation energy ratio $R_{J/2}\equiv E_x(J)/E_x(2^+)$, (b) normalized  staggering amplitude $S_{J/2}\equiv \Big\vert E_x(J)- \frac{J+1}{2J + 1} E_x(J-1)-\frac{J}{2J + 1} E_x(J+1)\Big\vert/E_x(2^+)$~\cite{Wiedenhover99} and (c) average deformation $\langle \beta_\lambda\rangle$ with $\lambda = 2, 3$ as functions of spin $J$ in $^{224}$Ra.}
\end{figure}

The calculated low-lying energy spectra are compared with available data in Fig.~\ref{fig3}. The spectra and the E2 and E3 transitions can be reproduced reasonably well using only the energy-minimum configuration (cf. Fig.~\ref{fig3}(c)). However, the predicted alternating parity rotation band as expected for a stable octupole shaped nucleus is not supported by the data. The energy displacement between the parity doublets can only be reproduced in the GCM calculation (cf. Fig.~\ref{fig3}(b)) by mixing the configurations around the equilibrium shape, which does not affect appreciably the good description for the E$\lambda$ strengths. As clearly shown in Fig.~\ref{fig4}, the gradually reduced odd-even staggering $R_{J/2}$ with increasing $J$ is associated with the similar behavior of average deformation parameters $\langle \beta_\lambda\rangle\equiv \sum_q \vert \beta_\lambda\vert \vert g^{J\pi}_\alpha\vert^2$, where the orthonormal collective wave function $g^{J\pi}_\alpha$ is constructed as~\cite{Ring80} $g^{J\pi}_\alpha (q_a) = \sum_{q_b} \left[\mathscr{N}^{J\pi}\right]^{1/2}_{q_a, q_b}f^{J\pi\alpha}_{q_b}$.
The octupole deformations of negative-parity states are almost constant (around 0.144) with increasing spin $J$, cf. Fig.~\ref{fig4}(c), while those of positive-parity states increase smoothly from $\langle\beta_3\rangle = 0.101$ ($J = 0$) to $\langle\beta_3\rangle =0.141$ ($J = 10~\hbar$). A similar but much less evident staggering is shown in $\langle\beta_2\rangle$ varying in between 0.180 and 0.191.

It is worth mentioning that in contrast to the predictions based on the energy-minimum configuration, the full configuration-mixing calculation predicts a steady increase (but also with somewhat odd-even staggering) of electric octupole transition moments $Q_3(J\to J-1)$ and $Q_3(J\to J-3)$. However, the available data on the E$\lambda$ transitions in $^{224}$Ra is limited and can hardly distinguish these two situations. Moreover, the ratio of electric dipole moment $Q_1(J\to J-1)$ to quadrupole moment $Q_2(J\to J-2)$ is similar in both calculations and not much dependent on spin $J$, which is consistent with the available data~\cite{Cocks97}.  In short, Fig.~\ref{fig4} presents us a novel spin-controlled shape stabilization picture that rotation brings the dominate shape of positive-parity (even-$J$) states close to that of negative-parity (odd-$J$) states. A further high-precise measurement on the $Q_3(J\to J-1)$ transition moments should be helpful to confirm this picture derived from the excitation energies.

\begin{figure}[tb]
\centering
\includegraphics[clip=,width=8.5cm]{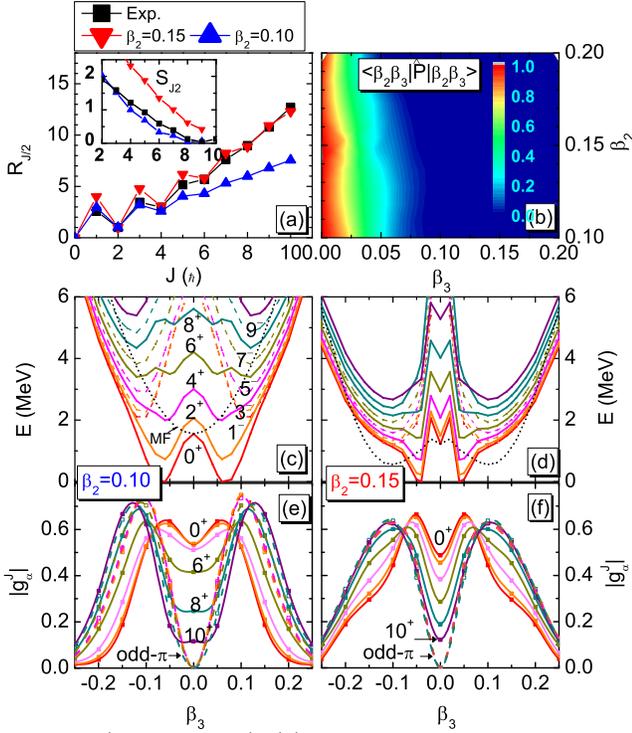}\vspace{-0.5cm}
\caption{\label{fig5} (Color online) (a) Excitation energy ratio $R_{J/2}$ and
 (inset) staggering amplitude $S_{J/2}$; (b) norm overlap $\langle \beta_2\beta_3\vert\hat P\vert  \beta_2\beta_3\rangle$ of parity operator in mean-field states; (c) and (d) mean-field and projected energy curves onto particle numbers $N, Z$, angular momentum $J$, and parity $\pi=(-1)^J$ (solid lines for even $J$, dashed lines for odd $J$). The mean-field (or projected) energy curves are normalized to the $0^+$ state at $\beta_3=0$ (or minimum). The mean-field energy surfaces in (c) and (d) are shifted down by 3.43 MeV and 4.15 MeV, respectively. (e) and (f) absolute value of collective wave functions for parity-doublet states. Those for odd-$\pi$ ($1^-, 3^-, \ldots, 9^-$) states are close to each other.}
\end{figure}

 To understand the connection between spin-dependent parity splitting and shape evolution in a simple and intuitive way, we carry out the GCM calculations by mixing the octupole configurations at either $\beta_2 = 0.10$ or 0.15.  Fig.~\ref{fig5}(a) displays the corresponding excitation energy ratio $R_{J/2}$ and normalized staggering amplitude $S_{J/2}$. The gradually reduced energy staggering is shown in both cases. The mechanism responsible for this behavior is exhibited in Fig.~\ref{fig5}(c) and (d) for the projected energy surfaces and in Fig.~\ref{fig5}(e) and (f) for the collective wave functions of each spin-parity states.  We note that the peak at $\beta_3=0.02$ in Fig.~\ref{fig5}(d) is due to the collapse of pairing correlation between protons and this configuration turns out to have a minor influence on the GCM results.

 In the projected GCM calculation, energy splitting between $\Delta J=1$ states comes mainly from the dynamical correlation energies (DCE) from angular-momentum and parity projections. The DCE from parity projection is decreasing to zero with increasing $\beta_3$ as the norm overlap $\langle \beta_2\beta_3\vert\hat P\vert  \beta_2\beta_3\rangle \to 0$, cf. Fig.~\ref{fig5}(b). Moreover, the larger the $\beta_2$ is, the faster the norm overlap approaches zero with $\beta_3$. Therefore, the energy splitting between $\Delta J=1$ states decreases with increasing of both $\beta_2$ and $\beta_3$, as shown in the projected energy surfaces, cf. Fig.~\ref{fig5}(c) and (d). An alternating-parity projected energy surfaces is shown in sufficient large deformation regions, where the energy splitting between $\Delta J=1$ states is originated almost purely from rotation, as expected for stable octupole shaped nuclei. In contrast, for the configurations in relatively small deformation regions, such as $\beta_3\lesssim0.05$, the nonzero overlap  $\langle \beta_2\beta_3\vert\hat P\vert  \beta_2\beta_3\rangle$ leads to a large energy splitting between $\Delta J=1$ parity doublets and thus to a large staggering amplitude.

 In particular, Fig.~\ref{fig5}(c) and (d) shows that the projected energy surfaces with positive parity evolve evidently with increasing spin, presenting a transition from weakly deformed octupole shape to a large deformed one, which is consistent with the behavior of the spin-dependent collective potentials introduced phenomenologically in Ref.~\cite{Jolos94}. This spin-controlled shape stabilization picture is demonstrated more clearly in the distributions of collective wave functions in Fig.~\ref{fig5}(e) and (f). With the increase of spin, the dominated configuration of positive-parity state drifts gradually from the weak octupole configurations ($\beta_3\simeq0.05$) to those with large octupole shapes ($\beta_3\in[0.10, 0.15]$). This behavior is probably related to the deformation-dependent moment of inertia, which turns out to increase with $\beta_2$ and $\beta_3$, presenting a peak region around the minimum of the mean-field energy surface in Fig.~\ref{fig2}. In contrast, due to antisymmetry requirement ($\beta_3 \to -\beta_3$) for the negative-parity states, their collective wave functions are zero at $\beta_3=0$, concentrated to a large octupole deformed configurations ($\beta_3\in[0.10, 0.15]$) and are hardly affected by rotation. It is coincident with the evolution picture of the collective wave functions  with spin from the full GCM calculation.

In conclusion, we have reported the first symmetry-conserved beyond-mean-field study of low-lying parity doublets states in $^{224}$Ra with a state-of-the-art multireference relativistic energy density functional method, where the dynamical correlations related to restoration of broken symmetries and to fluctuations of quadrupole-octupole shapes have been taken into account with the exact generator coordinate method combined with particle-number, angular-momentum, and parity projections. Both the energy spectrum and the E2, E3 transitions are reproduced in the configuration-mixing calculation without introducing any phenomenological parameter. We have demonstrated full microscopically that $^{224}$Ra has dynamical octupole shapes in the low-spin positive-parity states and acquires a stable octupole shape close to that of negative-parity states after rotation with spin up to $\sim8 \hbar$. This novel picture of rotation-induced octupole shape stabilization provides a natural explanation for the spin-dependent parity splitting in the excitation energies and is expected to be a common phenomenon in some actinides and rare-earth nuclei. Finally, we point out that with our microscopic results as the inputs of the coupled-channel calculation for nuclear fusion~\cite{Hagino2015}, one may observe some interesting differences on the barrier distribution compared with that based on rotational or vibrational limits~\cite{Kumar2015}. Moreover, octupole deformed odd-mass nuclei have enhanced time-reversal violating nuclear Schiff moments and thus are relevant for measuring atomic electric-dipole moments~\cite{Parker15}. The present study  provides a starting point to examine the dynamical correlation effects on nuclear  Schiff moments.

The authors thank J. Meng, T. Nik\u{s}i\'{c}, A. Vitturi, and D. Vretenar for helpful discussions, thank J. Engel and K. Hagino for careful reading the manuscript and thank J. L. Egido and L. M. Robledo for drawing our attention to their past works.  This work was supported partially by the NSFC under Grant Nos. 11575148, 11475140, 11305134, the Natural Science Foundation of Chongqing cstc2011jjA0376, and by the Chinese-Croatian project ``Nuclear Low-lying Spectrum and Related Hot Topics".

\end{document}